\documentclass[pss]{wiley2sp} % provides pss two-column style
\usepackage{amsmath}
\usepackage{bm}              % uncomment these two packages if you
\usepackage{color}

\begin{document}

% Title of the article
\title{Effect of Co doping and hydrostatic pressure on SrFe$_\mathbf{2}$As$_\mathbf{2}$}

% Abbreviated title for the page headers
\titlerunning{Effect of Co doping and hydrostatic pressure on SrFe$_2$As$_2$}

% Authors
\author{%
  E. Lengyel\textsuperscript{1},
  M. Kumar\textsuperscript{1,$\dag$},
  W. Schnelle\textsuperscript{1},
  A. Leithe-Jasper\textsuperscript{1}, and
  M. Nicklas\textsuperscript{1,$\ast$}
  }
% Abbreviated list of authors for the page headers
\authorrunning{E. Lengyel et al.}

%E-mail-address of corresponding author

\mail{e-mail
  \textsf{nicklas@cpfs.mpg.de}\\
\textsuperscript{$\dag$} Present address: Malaviya National Institute of Technology Jaipur, Department of Physics, Jaipur 302017, Rajasthan, India.%, Phone:
  %+xx-xx-xxxxxxx, Fax: +xx-xx-xxx
  }

% author's affiliations/addresses
\institute{%
  \textsuperscript{1}\,Max Planck Institute for Chemical Physics of Solids,
N\"{o}thnitzer Str.\ 40, 01187 Dresden, Germany.}

%\received{XXXX, revised XXXX, accepted XXXX} % do not change, will be filled in by the publisher
%\published{XXXX} % do not change, will be filled in by the publisher

% Please select about four verbal keywords for your manuscript.

%\dag

\abstract{%

\abstcol{%
We report a pressure study on electron doped SrFe$_{2-x}$Co$_x$As$_2$ by electrical-resistivity ($\rho$) and  magnet\-ic-susceptibility ($\chi$) experiments. Application of either external pressure or Co substitution rapidly suppresses the spin-density wave ordering of the Fe moments and induces superconductivity in SrFe$_2$As$_2$. At $x=0.2$ the broad superconducting (SC) dome in the $T-p$ phase diagram exhibits its maximum $T_{c,{\rm max}}=20$~K at a pressure of only $p_{\rm max}\approx 0.75$~GPa. In SrFe$_{1.5}$Co$_{0.5}$As$_2$ no superconductivity is observed anymore up to 2.8~GPa.}{Upon increasing the Co concentration the maximum of the SC dome shifts toward lower pressure accompanied by a decrease in the value of $T_{c,{\rm max}}$. Even though, superconductivity is induced by both tuning methods, Co substitution leads to a much more robust SC state. Our study evidences that in SrFe$_{2-x}$Co$_x$As$_2$ both, the effect of pressure and Co-substitution, have to be considered in order to understand the SC phase-diagram and further attests the close relationship of SrFe$_2$As$_2$ and its sister compound BaFe$_2$As$_2$.}}

\maketitle   % please do not remove

\section{Introduction}

The parent compounds \textit{A}Fe$_2$As$_2$ (\textit{A}=Ca, Sr, Ba, Eu) of the iron-pnictide superconductors  exhibit at ambient pressure a spin-density wave (SDW) instability at $T_{\rm SDW}$ which is associated with a tetragonal to orthorhombic structural phase transition \cite{Rotter08,Jesche08}. Doping with K, Cs, or Na at the \textit{A} site (hole doping) \cite{Johrendt09,sasmal,gchen,goko,Kasinathan09} or partial replacement of Fe with Co or Ni (electron doping) \cite{Kasinathan09,alj,saha,Sefat08} suppresses the SDW transition rapidly. Alternatively $T_{\rm SDW}$ can be suppressed by application of external pressure \cite{Kumar,Colombier09,miclea}. In both cases, superconductivity starts to develop in the region where $T_{\rm SDW}$ becomes zero. In SrFe$_2$As$_2$, for example, superconductivity has been confirmed on doping with K, Cs, or Na at the Sr-site (hole doping) \cite{Johrendt09,sasmal,gchen,goko,Kasinathan09} or partial replacement of Fe by Co or Ni (electron doping) \cite{Kasinathan09,alj,saha} as well as under external pressure  \cite{Kumar,Colombier09,alireza,Torikachvili08a,Matsubayashi09,Igawa09,Kotegawa09,Kitagawa09,Uhoya11,Wu14,Morozova15}.

While pressure-induced superconductivity in stoichiometric SrFe$_2$As$_2$ \cite{Kumar,Colombier09,alireza,Torikachvili08a,Matsubayashi09,Igawa09,Kotegawa09,Kitagawa09,Uhoya11,Wu14,Morozova15} as well as superconductivity on electron- and hole-doping has been studied extensively \cite{sasmal,gchen,goko,Kasinathan09,alj,saha,Schnelle09,Kim10,Muraba10,Weicker11}, only one combined pressure -- hole-doping study has been reported in Sr$_{1-x}$K$_x$Fe$_2$As$_2$ \cite{Gooch08}, but no pressure -- electron-doping investigation. Substituting Fe by Co in SrFe$_{2-x}$Co$_x$As$_2$ corresponds to an electron doping directly inside the iron-arsenide layers. Upon increasing the Co concentration the SDW transition is suppressed rapidly from $T_{\rm SDW}=205$~K in undoped SrFe$_2$As$_2$ to 95 K at a Co concentration of $x=0.15$ \cite{alj}. At $x=0.2$ no indication of the SDW anomaly is observed neither in resistivity nor in magnetic-susceptibility experiments \cite{alj}. Superconductivity develops in the concentration range $0.2\leq x\leq0.4$. The bulk nature of the superconductivity has been confirmed by specific-heat measurements \cite{alj}. The maximum $T_c$ of $T_{c,{\rm max}}=19.4$~K is already observed at $x=0.2$, which is the lowest Co concentration where superconductivity appears. With further increasing Co concentration $T_c$ decreases monotonically. Finally, SrFe$_{1.5}$Co$_{0.5}$As$_2$ is not superconducting (SC) anymore. The $T-x$ phase diagram of SrFe$_{2-x}$Co$_x$As$_2$ resembles that of the temperature -- pressure ($T-p$) phase diagram of stoichiometric SrFe$_2$As$_2$. Here, on application of hydrostatic pressure $T_{\rm SDW}$ is suppressed between 3 and 4~GPa \cite{Kumar}. In this pressure range also superconductivity appears \cite{Kumar}. The bulk nature of the SC phase has been confirmed by magnetization and susceptibility data \cite{alireza}. However, there is a discrepancy in literature on the actual pressure range where superconductivity is observed  \cite{Kumar,Colombier09,alireza,Torikachvili08a,Matsubayashi09,Igawa09,Kotegawa09,Kitagawa09,Uhoya11,Wu14,Morozova15}.

In this paper we compare the effect of electron doping inside the iron-arsenide layers by substituting Fe by Co and the application of hydrostatic pressure on the SDW transition as well as on the SC properties of SrFe$_2$As$_2$. While there is no data available on the combined effect of Co-substitution and pressure in SrFe$_2$As$_2$, its stoichiometric sister compound BaFe$_{2}$As$_2$ has been studied in details \cite{Ahilan08,Ahilan09,Colobier10,Drotziger10,Arsenijevic11,Zheng14,Tang14}. In order to establish the $T-p$ phase diagram of SrFe$_{2-x}$Co$_x$As$_2$ and to compare it with the findings in BaFe$_{2-x}$Co$_x$As$_2$, we carried out electrical-resistivity experiments under hydrostatic pressure on three different SrFe$_{2-x}$Co$_x$As$_2$ samples with the concentrations $x=0.1$, 0.2, and 0.5. The samples were chosen in a way to be in the underdoped ($x=0.1$, SDW and no superconductivity), optimally doped ($x=0.2$, no SDW and superconductivity), and overdoped ($x=0.5$, no SDW and no superconductivity at atmospheric pressure) regime. We will demonstrate that the
application of pressure gradually suppresses $T_{\rm SDW}$ and induces superconductivity in SrFe$_{1.9}$Co$_{0.1}$As$_2$, but at much lower pressures and with a much smaller $T_{c,{\rm max}}$ compared with stoichiometric SrFe$_2$As$_2$. For $x=0.2$ superconductivity appears in a dome-like shape in an extended pressure range. No superconductivity is induced in SrFe$_{1.5}$Co$_{0.5}$As$_2$. The electrical-resistivity studies are complemented by magnetic-susceptibility investigations on SrFe$_{1.85}$Co$_{0.15}$As$_2$ under pressure.

\section{Methods}

Polycrystalline samples were synthesized by sintering stoichiometric amounts of the precursors SrAs, Fe$_2$As and Co$_2$As (for details see Ref.\ \cite{alj}). X-ray diffraction measurements confirmed the ThCr$_2$Si$_2$-type structure (space group \textit{I}4/\textit{mmm}) for all samples \cite{alj}. Temperatures down to 1.8~K and magnetic fields up to 14~T were generated using a cryostat equipped with a superconducting magnet (PPMS, Quantum Design). In a double-layer piston-cylinder type cell pressures up to 2.8~GPa have been achieved with silicon oil as pressure transmitting medium \cite{Nicklas15}. A piece of lead was used as manometer. Measurements of the electrical resistivity were carried out using a standard four-probe technique utilizing a LR700 (Linear Research) resistance bridge. The magnetic field was applied perpendicular to the electrical current. Magnetic-susceptibility measurements were conducted with a miniaturized coil system consisting of one primary coil and one pair of compensated secondary coils sitting inside the pressure cell. Here the LR700 served as mutual-inductance bridge. A frequency of $\nu=16$~Hz and a modulation field, $H\rm _{AC}$, in the range between $0.06$ and $20$~Oe were used. The sample was placed in the center of one of the secondary coils.  In order to obtain absolute values of the sample susceptibility we used the size of the diamagnetic signal of the piece of lead sitting next to the sample and set it to $-1$. This procedure allowed us to estimate the SC volume fraction of our sample.

\section{Results and discussion}

%Electrical resistivity

\begin{figure}[t]%
\includegraphics*[width=\linewidth]{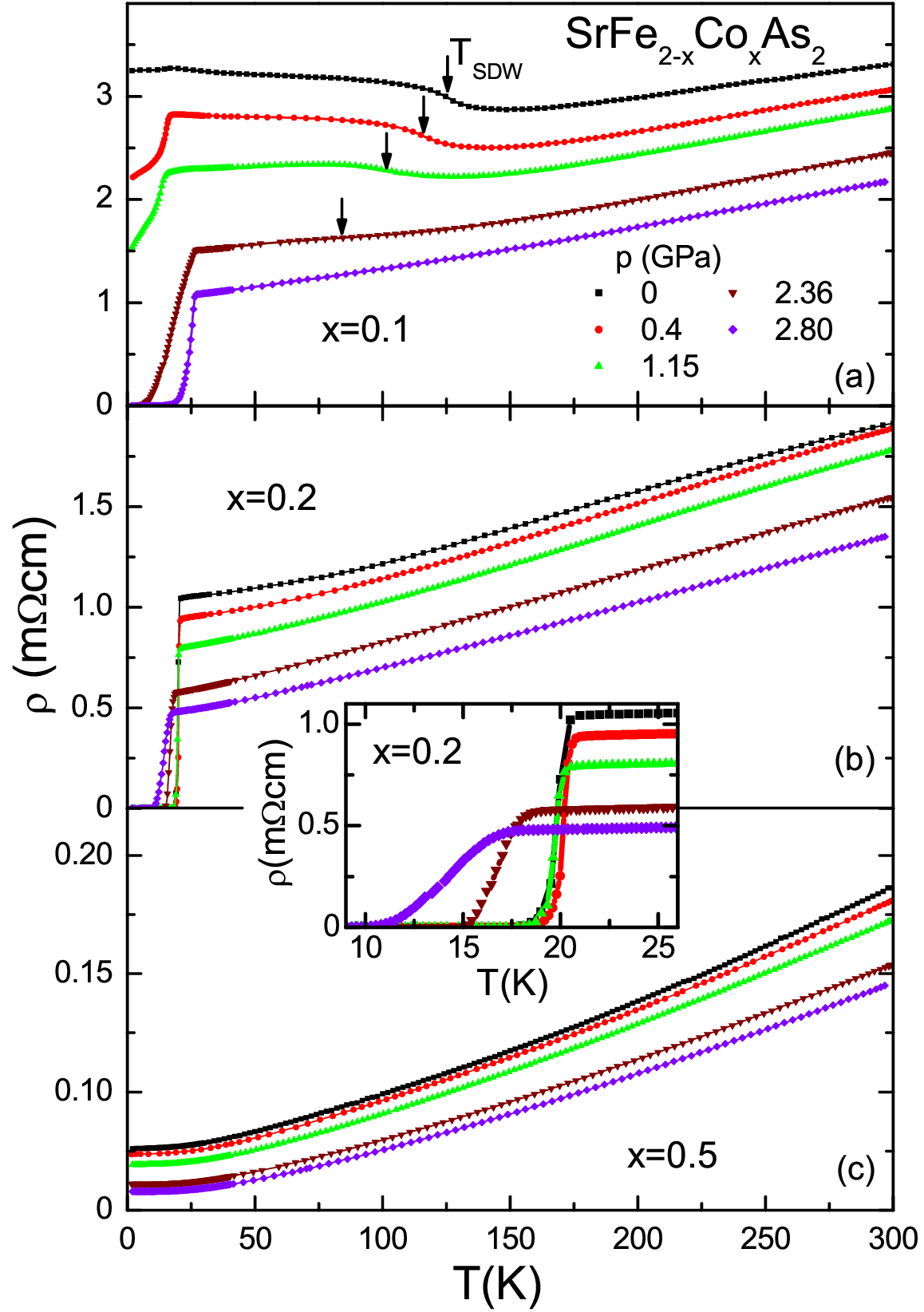}
\caption{%
 Electrical resistivity of SrFe$_{2-x}$Co$_x$As$_2$ ($x=0.1$, 0.2, and 0.5) for selected pressures. $T_{\rm SDW}$ is marked by arrows \cite{TSDW}. Inset: low-temperature electrical resistivity of SrFe$_{1.8}$Co$_{0.2}$As$_2$.}
\label{resistivity_low}\label{resistivity}
\end{figure}

Figure \ref{resistivity} displays the electrical resistivity $\rho(T)$ of SrFe$_{2-x}$Co$_x$As$_2$
($x=0.1$, 0.2, and 0.5) for selected pressures. At all investigated pressures, $\rho(T)$ decreases
monotonically upon decreasing temperature indicating metallic behavior. Only for $x=0.1$ a pronounced
upturn in $\rho(T)$ toward lower temperatures indicates the SDW transition of the itinerant iron moments, which is associated with a structural transition from a tetragonal to an orthorhombic phase. The increase in $\rho(T)$ at $T_{\rm SDW}$ \cite{TSDW}, is distinct from the sharp drop observed in undoped SrFe$_2$As$_2$ \cite{Jesche08}, but is in good agreement with reports on other doped \textit{A}Fe$_2$As$_2$ compounds \cite{alj,Ahilan08,Torikachvili08,Nicklas10}. The different behavior in the electrical resistivity at $T_{\rm SDW}$ might be related to disorder introduced by the Co substitution. In SrFe$_{1.9}$Co$_{0.1}$As$_2$ application of pressure shifts the anomaly at the SDW transition from $T_{\rm SDW}=125$~K at ambient pressure to 72~K at 2.60~GPa. The anomaly in $\rho(T)$ broadens and becomes less pronounced with increasing pressure. At 2.8~GPa we do not find any feature related to the SDW transition in our data anymore suggesting that $T_{\rm SDW}(p)$ is suppressed to zero at a critical pressure $p_c$ between 2.6 and 2.8~GPa. The resistivity curves of SrFe$_{2-x}$Co$_x$As$_2$, $x=0.2$ and 0.5, do not show any qualitative change in shape upon increasing pressure (see Fig.\ \ref{resistivity}). While SrFe$_{1.9}$Co$_{0.1}$As$_2$ and SrFe$_{1.8}$Co$_{0.2}$As$_2$ become superconducting, SrFe$_{1.5}$Co$_{0.5}$As$_2$ does not show any indication of the formation of a SC state in the whole investigated pressure range up to $2.8$~GPa.

\begin{figure}[t]%
\includegraphics*[width=\linewidth]{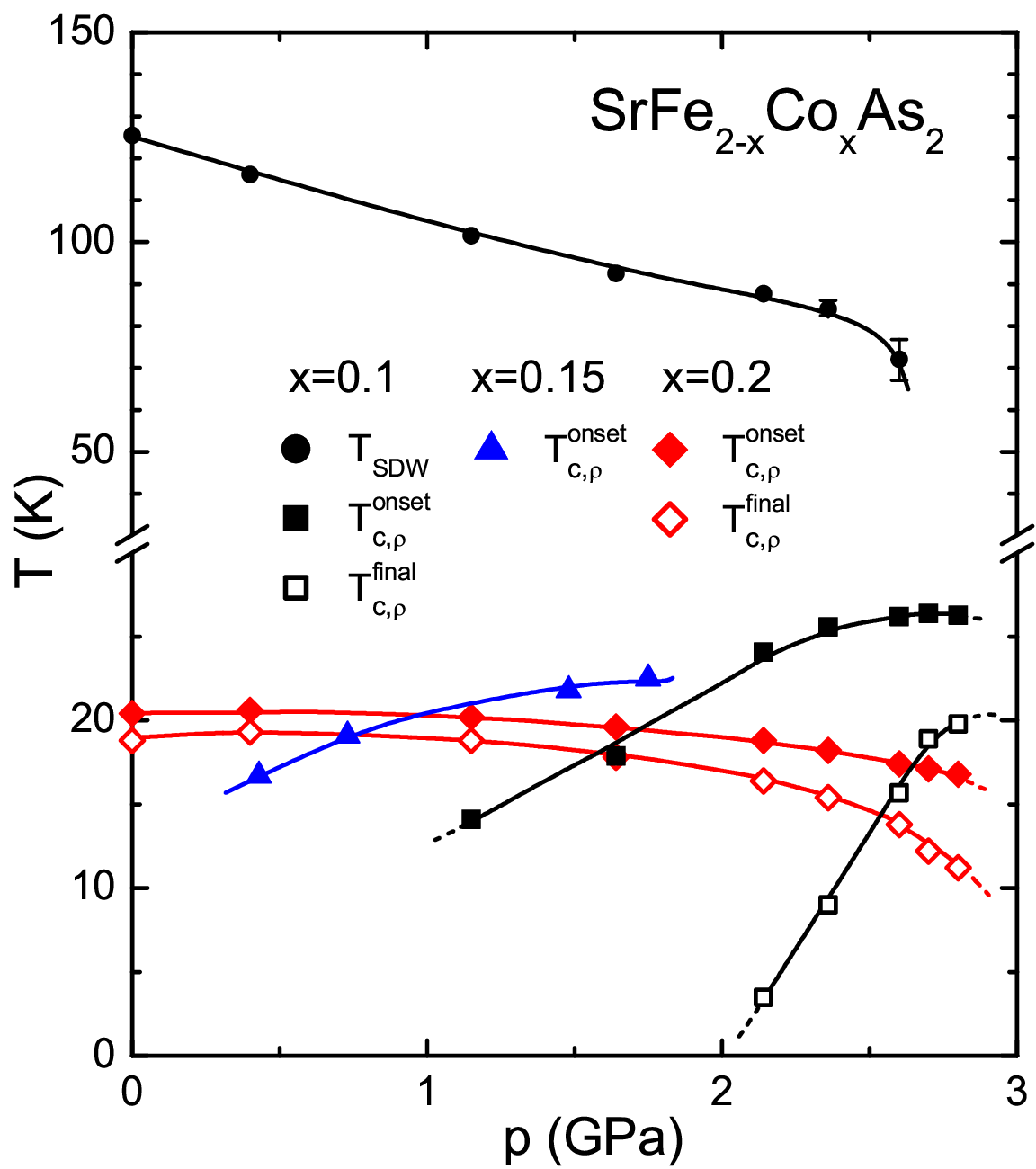}
\caption{%
 $T-p$ phase diagram of SrFe$_{2-x}$Co$_x$As$_2$, $x=0.1$, 0.15 (resistivity data not shown), and 0.2. $T_{\rm SDW}$, $T^{\rm onset}_{c,\rho}$, and $T^{\rm final}_{c,\rho}$ obtained from the electrical resistivity as function of pressure \cite{note_rho}.}
\label{PhaseDiagram}
\end{figure}

In SrFe$_{1.9}$Co$_{0.1}$As$_2$ a small reduction in $\rho(T)$ below 15~K indicates the presence of some spurious superconductivity already at ambient pressure (see Fig.~\ref{resistivity}a), which might be induced by internal strains \cite{saha}. Upon increasing pressure the drop in $\rho(T)$ becomes more pronounced and its onset shifts to higher temperatures. Finally, at $p=2.36$~GPa a zero-resistance state is observed below $T^{\rm final}_{c,\rho}= 4.0$~K \cite{note_rho}. We note that at this pressure the SDW transition is still present at $T_{\rm SDW}=84$~K along with superconductivity at low temperatures. The step in $\rho(T)$ becomes sharper and  $T^{\rm final}_{c,\rho}$ increases up to $19.8$~K at 2.8~GPa. The increase in $T_c$ coincides with a considerable narrowing of the transition anomaly in the resistivity, as shown in the $T-p$ phase diagram in Fig.\ \ref{PhaseDiagram}. The width of the SC transition decreases upon increasing pressure from $\Delta T_c=T^{\rm onset}_{c,\rho}-T^{\rm final}_{c,\rho}=20.6$~K at 2.14~GPa to only 6.5~K at 2.8~GPa. While $T^{\rm final}_{c,\rho}$ still increases slightly up to our highest pressure $p=2.8$~GPa, $T^{\rm onset}_{c,\rho}$ achieves its maximum already at 2.7~GPa denoting that the maximum of the SC dome is almost reached. This indicates that in SrFe$_{1.9}$Co$_{0.1}$As$_2$ $T_{c,{\rm max}}$ is smaller than in the undoped mother compound \cite{Sefat11}. One reason might be the disorder induced in the iron-arsenide layers by the Co substitution which could limit $T_c$ under pressure.

We now turn to SrFe$_{1.8}$Co$_{0.2}$As$_2$, where at ambient pressure no SDW order is present anymore. A drop in $\rho(T)$ to zero at $T^{\rm final}_{c,\rho}=18.8$~K indicates the existence of superconductivity already at atmospheric pressure. The bulk nature of the SC phase has been confirmed by magnetic-susceptibility and specific-heat experiments \cite{alj}. The effect of pressure on $T_c$ is initially rather weak. The inset of Fig.\ \ref{resistivity} displays a magnification of the resistivity in the low-temperature region. Upon application of pressure the SC transition temperature first increases slightly up to $T_{c,{\rm max}}=19.3$~K at $0.4$~GPa before it starts to decrease again, indicating that SrFe$_{1.8}$Co$_{0.2}$As$_2$ is still situated on the left side of the SC dome, in the under-doped regime, in the $T-x$ phase diagram. Above $1.5$~GPa, $T^{\rm final}_{c,\rho}(p)$ starts to drop faster upon increasing pressure as depicted in Fig.\ \ref{PhaseDiagram}. In the region of the maximum of the SC dome we observe a rather narrow SC transition,  $\Delta T_c=1.3$~K. This value is considerably smaller than in the case of SrFe$_{1.9}$Co$_{0.1}$As$_2$ despite the higher Co concentration. This might indicate that disorder induced by Co substitution has only a minor effect on the width of the SC transition. With increasing distance to $T_{c,{\rm max}}$ $\Delta T_c$ increases slightly, but still remains much smaller than in SrFe$_{1.9}$Co$_{0.1}$As$_2$.

The effect of pressure on SrFe$_{2}$As$_2$ resembles that of Co-substitution on the iron site \cite{alj,Kumar}. However, the maximum SC transition temperature is larger in pressurized compared with Co-doped SrFe$_{2}$As$_2$. Our study shows that this is a general trend in the combined $T-p-x$ phase diagram of SrFe$_{2-x}$Co$_x$As$_2$: $T_{c,{\rm max}}$ observed under pressure decreases and shifts toward lower temperatures upon increasing the Co concentration. At the same time superconductivity develops in an extended pressure interval. A similar behavior has been previously established for BaFe$_{2-x}$Co$_x$As$_2$ \cite{Ahilan08,Ahilan09,Colobier10,Drotziger10,Arsenijevic11,Zheng14,Tang14}. Thus our study provides additional experimental evidence for the close relation of the Sr- and Ba-variants of the 122 iron-pnictide superconductors.

\begin{figure}[t]%
\includegraphics*[width=\linewidth]{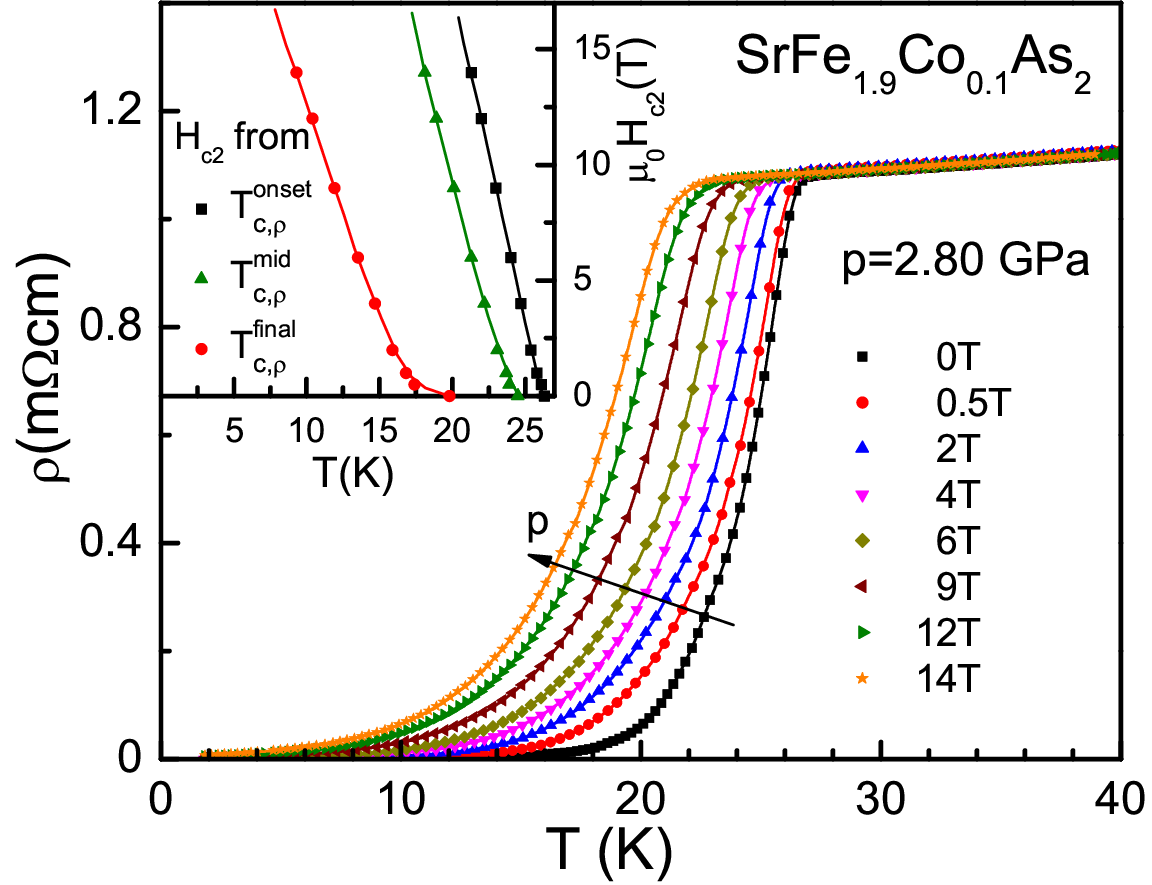}
\caption{%
  Electrical resistivity of SrFe$_{1.9}$Co$_{0.1}$As$_2$ at 2.8 GPa in different applied magnetic fields. Inset: $H_{c2}-T$ diagram for $H_{c2}$ determined from $T^{\rm onset}_{c,\rho}$, $T^{\rm mid}_{c,\rho}$, and $T^{\rm final}_{c,\rho}$ \cite{note_rho}.}
\label{resistivityField}
\end{figure}

In SrFe$_{1.9}$Co$_{0.1}$As$_2$ we have also investigated the effect of a magnetic field on the SC transition at 2.8~GPa close to the maximum of the SC dome. The results of the $\rho(T)$ experiments  carried out in different magnetic fields up to 14~T are displayed in Fig.\ \ref{resistivityField}. We note that the transition slightly broadens upon increasing the magnetic field. This is expected in a polycrystalline material possessing an anisotropic $H_{c2}(T)$ as anticipated for a quasi-two dimensional electronic structure \cite{Weicker11}. The $H_{c2}(T)$ curves obtained from $T^{\rm onset}_{c,\rho}$, $T^{\rm mid}_{c,\rho}$, and $T^{\rm final}_{c,\rho}$ are displayed in the inset of Fig.~\ref{resistivityField}. A minor tail in $H_{c2}(T)$ is visible at small magnetic fields. It is most pronounced in the $H^{\rm final}_{c2}(T)$ curve. It is likely to be related with multiband effects. Disregarding this tail, we obtain an initial slope of  ${\mu_0\rm d}H^{\rm final}_{c2}(T)/{\rm d}T \approx {-1.8}$~T/K.

%Magnetic susceptibility

\begin{figure}[t]%
\includegraphics*[width=\linewidth]{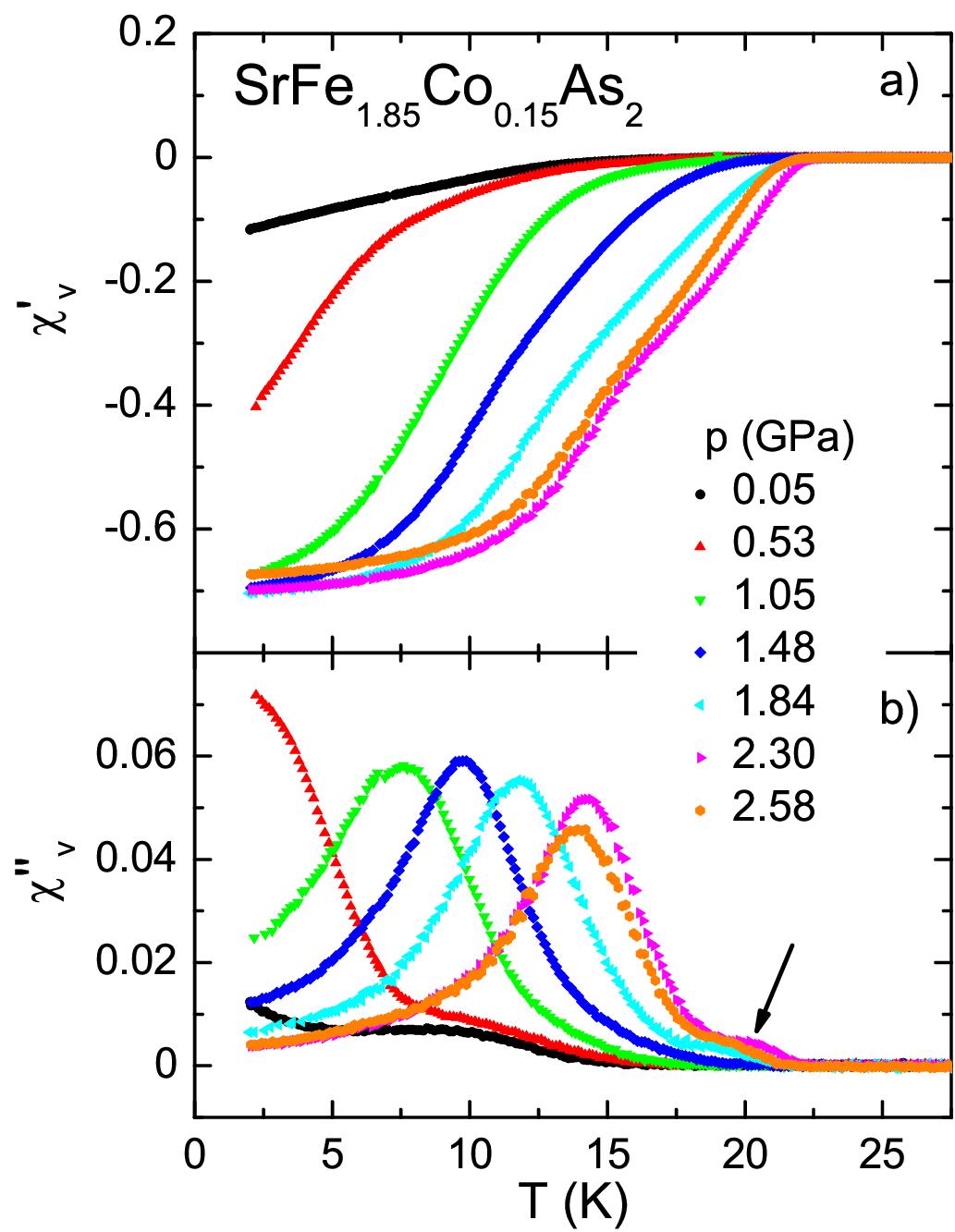}
\caption{%
  (a) Real and (b) imaginary parts of the volume susceptibility of SrFe$_{1.85}$Co$_{0.15}$As$_2$ at different pressures. The data were taken with an excitation frequency of 16~Hz and an amplitude of the oscillation field of 1.33~Oe. The arrow in (b) marks the shoulder in $\chi''_V(T)$ attributed to intragrain superconductivity. See text for details.}
\label{ChiTp}
\end{figure}

In addition to the electrical-resistivity studies, we carried out detailed magnetic-susceptibility measurements on SrFe$_{1.85}$Co$_{0.15}$As$_2$ under external pressure. Figure \ref{ChiTp} depicts the temperature dependence of the real $\chi'_V$ and imaginary $\chi''_V$ parts of the volume susceptibility at different pressures. $\chi'_V(T)$ at 0.05~GPa already displays a small diamagnetic response of about 10\% of the sample volume at 1.8~K. Upon increasing pressure the SC volume fraction rapidly increases. At 1.05~GPa $\chi'_V(T)$ tends to saturate toward the lowest temperatures at about $-0.7$ indicating a diamagnetic signal corresponding to 70~\% of the sample volume.  Within the accuracy of our experiment this value does not change up to $p=2.58$~GPa, the highest pressure in our experiment. Considering the uncertainties in the estimation of the SC volume fraction, our results indicate that almost the complete sample becomes SC in the pressure range above 1.05~GPa.

\begin{figure}[t]%
\includegraphics*[width=\linewidth]{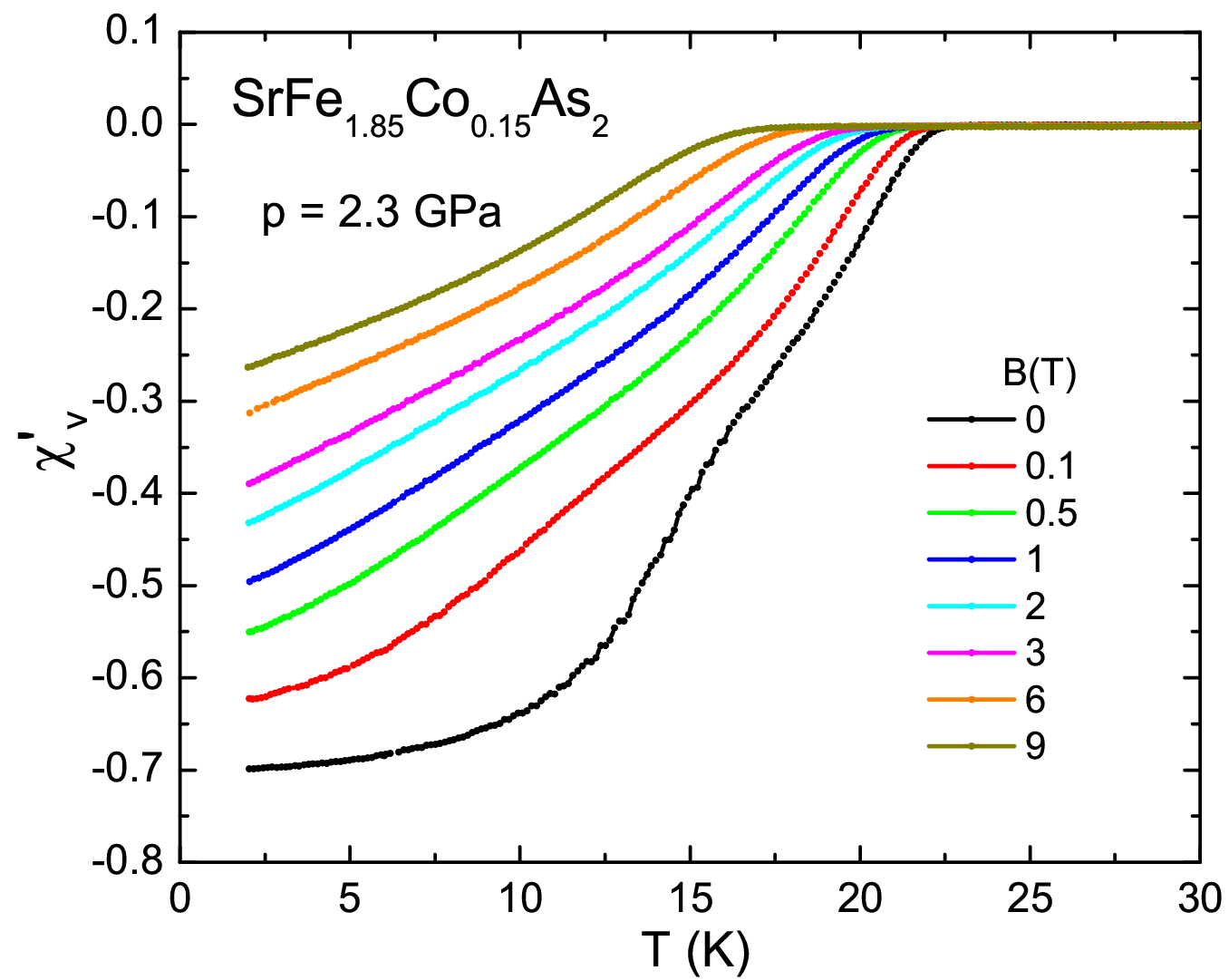}
\caption{%
  Real part of the volume susceptibility of SrFe$_{1.85}$Co$_{0.15}$As$_2$ at 2.3~GPa measured in different magnetic fields upon warming, after cooling the sample in zero magnetic field. The data were taken with an excitation frequency of 16~Hz and an amplitude of the oscillation field of 1.33~Oe.}
\label{Chi_field}
\end{figure}

The effect of a magnetic field on $\chi'_V(T)$ at 2.3~GPa, close to the maximum of the SC dome, is shown in Fig.\ \ref{Chi_field}. The SC volume fraction decreases fast upon increasing the magnetic field and does not saturate anymore down to the lowest temperatures already in the smallest applied field of 0.1~T. At 9~T we find a SC volume fraction of only 25\% remaining.

\begin{figure}[t]%
\includegraphics*[width=\linewidth]{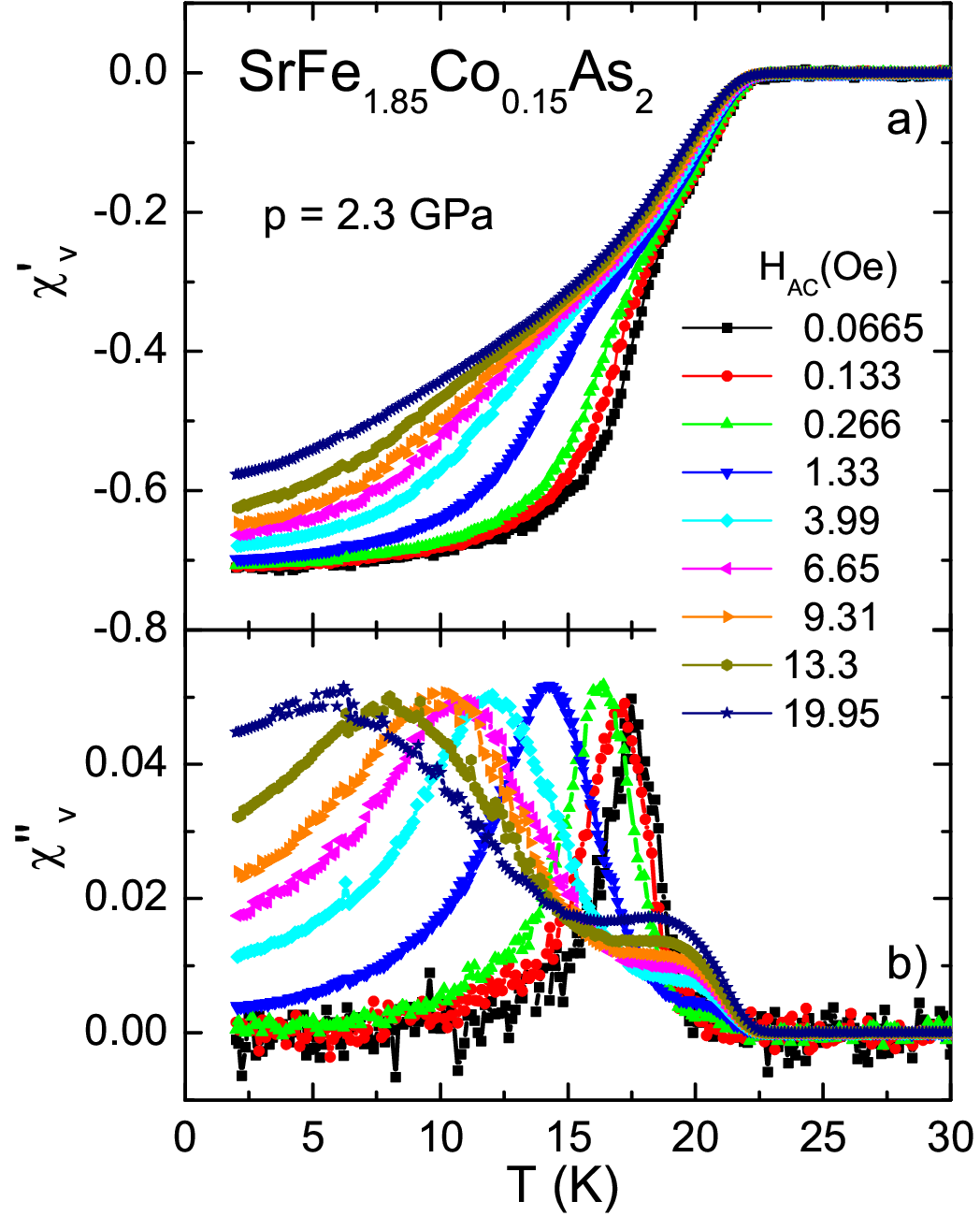}
\caption{%
  (a) Real and (b) imaginary parts of the volume susceptibility of SrFe$_{1.85}$Co$_{0.15}$As$_2$ at 2.3~GPa measured in zero magnetic field with different amplitudes of the oscillation field $H_{AC}$. The data were taken with an excitation frequency of 16~Hz.}
\label{Chi__AC_field}
\end{figure}

At 2.3~GPa, $\chi'_V(T)$ exhibits a sharp drop at the onset of the diamagnetic response, but develops a shoulder at lower temperatures. This shoulder is also present at smaller pressures, but less prominent. This two-step SC transition is known for polycrystalline samples and is related to the granularity of the material.  It has been investigated in details in the high-temperature cuprate superconductors \cite{Mueller89}. Upon cooling first intra-grain superconductivity develops before inter-grain coherence is established at lower temperatures. $\chi''_V(T)$ displays a large peak related to loss due to the coupling between the grains, while only a small feature is visible at the onset of the diamagnetic response arising from flux entry into the grains (see  Fig.\ \ref{ChiTp}). As expected, the amplitude of the oscillation field has a strong influence on inter-grain effects as visible in the magnetic susceptibility displayed in Fig.\ \ref{Chi__AC_field}. Upon changing the amplitude from $H_{\rm AC}=0.0665$ to 19.95~Oe by a factor of 300, the onset of the diamagnetic signal is almost unaltered. There is only little influence on the initial drop in $\chi'_V(T)$, since this drop originates from the development of superconductivity in individual grains. However, we find large changes due to inter-grain effects in both, real and imaginary parts of the susceptibility: increasing $H_{\rm AC}$ reduces the diamagnetic signal and shifts the inter-grain loss peak strongly toward lower temperatures. At the same time a second peak (shoulder) at higher temperatures associated with the flux entry into the grains becomes more pronounced.

\section{Summary}
We carried out a combined Co-substi\-tution and hydrostatic-pressure study for SrFe$_{2}$As$_2$ materials, where substituting Fe by Co directly corresponds to electron doping into the iron-arsenide layers. Co substitution as well as application of pressure leads to a suppression of the SDW transition and the development of a SC phase in a similar way like in BaFe$_{2}$As$_2$ . We observe a SC dome in the $T-p$ phase diagram for SrFe$_{2-x}$Co$_x$As$_2$, $x=0.1$, 0.15, and 0.2. Its maximum shifts systematically toward lower pressures with increasing Co concentration. At the same time the maximum $T_c$ decreases, but superconductivity extends to a larger pressure range in the $T-p$ phase diagram indicating a more robust SC state.

\section{Acknowledgements}
We thank the Deutsche For\-schungs\-gemeinschaft (DFG) for financial support through SPP~1458.


\begin{thebibliography}{[1]}

\bibitem{Rotter08} M. Rotter, M. Tegel, D. Johrendt, I. Schellenberg, W. Hermes, and R. P\"{o}ttgen, Phys. Rev. B \textbf{78}, 020503(R) (2008).

\bibitem{Jesche08} A. Jesche, N. Caroca-Canales, H. Rosner, H. Borrmann, A. Ormeci, D. Kasinathan, H. H. Klauss, H. Luetkens, R. Khasanov, A. Amato, A. Hoser, K. Kaneko, C. Krellner, and C. Geibel, Phys. Rev. B \textbf{78}, 180504(R) (2008).

\bibitem{Johrendt09} M. Rotter, M. Tegel, and D. Johrendt, Phys. Rev. Lett. 101, 107006 (2008).

\bibitem{sasmal} K. Sasmal, B. Lu, B. Lorenz, A. M. Guloy, F. Chen, Y. Xue, and C. Chu, Phys. Rev. Lett. {\bf 101}, 107007 (2008).

\bibitem{gchen}G. F. Chen, Z. Li, G. Li, W. Z. Hu, J. Dong, X. D. Zhang, P. Zheng, N. L. Wang, and J. L. Luo, Chin. Phys. Lett. {\bf 25} 3403 (2008).

\bibitem{goko} T. Goko, A. A. Aczel, E. Baggio-Saitovitch, S. L. Bud'ko, P.C. Canfield, J. P. Carlo, G. F. Chen, Pengcheng Dai, A. C. Hamann, W. Z. Hu, H. Kageyama, G. M. Luke, J. L. Luo, B. Nachumi, N. Ni, D. Reznik, D. R.
 Sanchez-Candela, A. T. Savici, K. J. Sikes, N. L. Wang, C. R. Wiebe, T. J. Williams, T. Yamamoto, W. Yu, and Y. J. Uemura, Phys. Rev. B \textbf{80}, 024508 (2009).

\bibitem{Kasinathan09}D. Kasinathan, A. Ormeci, K. Koch, U. Burkhardt, W. Schnelle, A. Leithe-Jasper, H. Rosner, New J. Phys. {\bf 11}, 025023 (2009).

\bibitem{alj}A. Leithe-Jasper, W. Schnelle, C. Geibel, and H. Rosner, Phys. Rev. Lett. {\bf 101} 207004 (2008).

\bibitem{saha} S. R. Saha, N. P. Butch, K. Kirshenbaum, and J. Paglione, Phys. Rev. B \textbf{79}, 224519 (2009).

\bibitem{Sefat08} A. S. Sefat, R. Jin, M. A. McGuire, B. C. Sales, D. J. Singh, and D. Mandrus, Phys. Rev. Lett. \textbf{101}, 117004 (2008).

\bibitem{Kumar} M. Kumar, M. Nicklas, A. Jesche, N. Caroca-Canales, M. Schmitt, M. Hanfland, D. Kasinathan, U. Schwarz, H. Rosner, and C. Geibel, Phys. Rev. B {\bf 78}, 184516 (2008).

\bibitem{Colombier09} E. Colombier, S. L. Bud'ko, N. Ni, and P. C. Canfield, Phys. Rev. B \textbf{79}, 224518 (2009).

\bibitem{miclea} C. F. Miclea, M. Nicklas, H. S. Jeevan, D. Kasinathan, Z. Hossain, H. Rosner, P. Gegenwart, C. Geibel, and F. Steglich, Phys. Rev. B {\bf 79}, 212509 (2009).

\bibitem{alireza} P. L. Alireza, Y. T. C. Ko, J. Gillett, C. M. Petrone, J. M. Coole, G. G. Lonzarich and S. E. Sebastian, J. Phys.: Conndens. Matter {\bf 21} 012208 (2008).

\bibitem{Torikachvili08a} M. S. Torikachvili, S. L. Bud'ko, N. Ni, and P. C. Canfield, Phys. Rev. B \textbf{78}, 104527 (2008).

\bibitem{Matsubayashi09} K. Matsubayashi, N. Katayama, K. Ohgushi, A. Yamada, K. Munakata, T. Matsumoto, and Y. Uwatoko, J. Phys. Soc. Jpn. \textbf{78}, 073706 (2009).

\bibitem{Igawa09} K. Igawa, H. Okada, H. Takahashi, S. Matsuishi, Y. Kamihara, M. Hirano, H. Hosono, K. Matsubayashi, and Y. Uwatoko, J. Phys. Soc. Jpn. \textbf{78}, 025001 (2009).

\bibitem{Kotegawa09} H. Kotegawa, H. Sugawara, and H. Tou, J. Phys. Soc. Jpn. \textbf{78}, 013709 (2009).

\bibitem{Kitagawa09} K. Kitagawa, N. Katayama, H. Gotou, T. Yagi, K. Ohgushi, T. Matsumoto, Y. Uwatoko, and M. Takigawa, Phys. Rev. Lett. \textbf{103}, 257002 (2009).

\bibitem{Uhoya11}  W. O. Uhoya, J. M. Montgomery, G. M. Tsoi, Y. K. Vohra, M. A. McGuire, A. S. Sefat, B. C. Sales, and S. T. Weir, J. Phys.: Condens. Matter {\bf 23}, 122201 (2011).

\bibitem{Wu14} J. J. Wu, J. F. Lin, X. C. Wang, Q. Q. Liu, J. L. Zhu, Y. M. Xiao, P. Chow, and C. Q. Jin, Sci. Rep. \textbf{4}, 3685 (2014).

\bibitem{Morozova15}  N. V. Morozova, A. E. Karkin, S. V. Ovsyannikov, Y. A. Umerova, V. V. Shchennikov, R. Mittal, and A. Thamizhavel, Supercond. Sci. Technol. \textbf{28}, 125010 (2015).

\bibitem{Schnelle09} W. Schnelle, A. Leithe-Jasper, R. Gumeniuk, U. Burkhardt, D. Kasinathan, and H. Rosner, Phys. Rev. B \textbf{79}, 214516 (2009).

\bibitem{Kim10} J. S. Kim, S. Khim, H. J. Kim, M. J. Eom, J. M. Law, R. K. Kremer, J. H. Shim, and K. H. Kim, Phys. Rev. B \textbf{82}, 024510 (2010).

\bibitem{Muraba10} Y. Muraba, S. Matsuishi, S.-W. Kim, T. Atou, O. Fukunaga, and H. Hosono, Phys. Rev. B \textbf{82}, 180512(R) (2010).

\bibitem{Weicker11} F. Weickert, M. Nicklas, W. Schnelle, J. Wosnitza, A. Leithe-Jasper, and H. Rosner, J. Appl. Phys. \textbf{110}, 123906 (2011).

\bibitem{Gooch08} M. Gooch, B. Lv, B. Lorenz, A. M. Guloy, and C. W. Chu, Phys. Rev. B {\bf 78}, 180508(R) (2008).

\bibitem{Ahilan08} K. Ahilan, J. Balasubramaniam, F. L. Ning, T. Imai, A. S. Sefat, R. Jin, M. A. McGuire, B. C. Sales, and D. Mandrus, J. Phys.: Condens. Matter {\bf20}, 472201 (2008).

\bibitem{Ahilan09} K. Ahilan, F. L. Ning, T. Imai, A. S. Sefat, M. A. McGuire, B. C. Sales, and D. Mandrus, Phys. Rev. B {\bf79}, 214520 (2009).

\bibitem{Colobier10} E. Colombier, M. S. Torikachvili, N. Ni, A. Thaler, S. L. Bud'ko, P. C. Canfield, Supercond. Sci. Technol. {\bf 23}, 054003 (2010).

\bibitem{Drotziger10} S. Drotziger, P. Schweiss, K. Grube, T. Wolf, P. Adelmann, C. Meingast, H. v. L\"{o}hneysen, J. Phys. Soc. Jpn. {\bf 79}, 124705 (2010).

\bibitem{Arsenijevic11} S. Arsenijevi\'{c}, R. Ga\'{a}l, A. S. Sefat, M. A. McGuire, B. C. Sales, D. Mandrus, L. Forr\'{o}, Phys. Rev. B {\bf84}, 075148, (2011).

\bibitem{Zheng14} Y. Zheng, Y. Wang, F. Hardy, A. E. B\"{o}hmer, T. Wolf, C. Meingast, and R. Lortz, Phys. Rev. B {\bf 89}, 054514 (2014).

\bibitem{Tang14} Y. Tang, Q, Tao, Z.A. Xu, X.J. Chen, J. Appl. Phys. {\bf 115}, 143904 (2014).

\bibitem{Nicklas15} M. Nicklas, in: Strongly Correlated Systems - Experimental Techniques, edited by A. Avella and F. Mancini (eds.), Springer Series in Solid-State Sciences 180, (Springer, Berlin, Heidelberg, 2015), pp.\ 173-204.

\bibitem{TSDW}$T_{\rm SDW}$ is defined by the position of the inflection point in $\rho(T)$.

\bibitem{Torikachvili08} M. S. Torikachvili, S. L. Bud'ko, N. Ni, and P. C. Canfield, Phys. Rev. Lett. {\bf 101} 057006 (2008).

\bibitem{Nicklas10} M. Nicklas, M. Kumar, E. Lengyel, W. Schnelle and A. Leithe-Jasper, J. Phys. Conf. Ser. {\bf273}, 012101 (2011).

\bibitem{note_rho} $T^{\rm onset}_{c,\rho}$, $T^{\rm mid}_{c,\rho}$, and $T^{\rm final}_{c,\rho}$ are defined by the temperature where $\rho(T)$ reaches 95\%, 50\%, and 5\% of the normal state resistivity, respecively.

\bibitem{Mueller89} K.-H. M\"{u}ller, Physica C \textbf{159}, 717 (1989).



\end{thebibliography}
\end{document}